\documentclass[a4paper,twoside,12pt]{article}
\input{obsjkt.sty}

\newcommand{\Msun}{\ensuremath{~{\rm M}_\odot}}                   
\newcommand{\Rsun}{\ensuremath{~{\rm R}_\odot}}                   
\newcommand{\rhosun}{\ensuremath{~\rho_\odot}}                    
\newcommand{\Teff}{\ensuremath{T_{\rm eff}}}                      
\newcommand{\EBV}{\ensuremath{E(B\!-\!V)}}                        
\newcommand{\degr}{\ensuremath{^\circ}}                           
\renewcommand{\kms}{~km~s$^{-1}$}                                 
\renewcommand{\cd}{~d$^{-1}$}                                     
\newcommand{\hip}{\textit{Hipparcos}}                             
\newcommand{\gaia}{\textit{Gaia}}                                 
\newcommand{\targ}{V596~Pup}
\newcommand{\targfull}{V596~Puppis}

\newcommand{\Msunnom}{\hbox{$\mathcal{M}^{\rm N}_\odot$}}
\newcommand{\Rsunnom}{\hbox{$\mathcal{R}^{\rm N}_\odot$}}
\newcommand{\Lsunnom}{\hbox{$\mathcal{L}^{\rm N}_\odot$}}

\usepackage{rotating}

\begin{document} 

\OBSheader{Rediscussion of eclipsing binaries: \targ}{J.\ Southworth}{2025 June}

\OBStitle{Rediscussion of eclipsing binaries. Paper XXIV. \\ The $\delta$~Scuti pulsator V596~Pup (formerly known as VV~Pyx)}

\OBSauth{John Southworth}

\OBSinstone{Astrophysics Group, Keele University, Staffordshire, ST5 5BG, UK}


\OBSabstract{\targ\ is a detached eclipsing binary containing two A1~V stars in a 4.596-d period orbit with a small eccentricity and apsidal motion, previously designated as VV~Pyxidis. We use new light curves from the Transiting Exoplanet Survey Satellite (TESS) and published radial velocities to determine the physical properties of the component stars. We find masses of $2.098 \pm 0.021$\Msun\ and $2.091 \pm 0.018$\Msun, and radii of $2.179 \pm 0.008$\Rsun\ and $2.139 \pm 0.007$\Rsun. The measured distance to the system is affected by the light from a nearby companion star; we obtain $178.4 \pm 2.5$~pc. The properties of the system are best matched by theoretical predictions for a subsolar metallicity of $Z=0.010$ and an age of 570~Myr. We measure seven significant pulsation frequencies from the light curve, six of which are consistent with $\delta$~Scuti pulsations and one of which is likely of slowly-pulsating B-star type.}


\section*{Introduction}

Eclipsing binary systems contain the only stars for which a direct measurement of their mass and radius is possible. Detached eclipsing binaries (dEBs) are an important class of these objects because their components have evolved as single stars. Their measured properties can be compared to the predictions of theoretical models of stellar evolution, to check the reliability of the predictions and help calibrate the physical ingredients of the models \cite{HiglWeiss17aa,ClaretTorres18apj,Tkachenko+20aa}.

Another approach to improving the theoretical descriptions of stars is that of asteroseismology \cite{Aerts++10book}. The measured stellar oscillation frequencies can be compared to theoretical predictions to infer their densities, ages, rotational profiles and the strength of chemical mixing \cite{Aerts+03sci,Briquet+07mn,Garcia+13aa,Bedding+20nat}.

Some dEBs show the signs of pulsations in one or both components, and thus might provide more exacting constraints on stellar theory. The pulsational types so far found in dEBs include $\delta$\,Scuti \citep{Hambleton+13mn,Maceroni+14aa,Me++23mn,Jennings+24mn2}, $\gamma$\,Doradus \cite{Debosscher+13aa,MeVanreeth22mn}, slowly-pulsating B-type (SPB) \cite{Clausen96aa,MeBowman22mn} and $\beta$~Cephei \cite{Me+20mn,Me++21mn,EzeHandler24apjs}. Several of these have been studied in previous papers in the current series: the hybrid $\delta$\,Sct/$\gamma$\,Dor systems RR~Lyn \cite{Me21obs6} and GK~Dra \cite{Me24obs1}, and the high-mass pulsators V1388~Ori \cite{Me22obs4} and V1765~Cyg \cite{Me23obs6}.

In this work we present an analysis of \targfull\ (Table~\ref{tab:info}) based on published spectroscopy and new space-based photometry. We include an independent discovery of $\delta$~Scuti pulsations in this object.


\section*{\targfull}

\begin{table}[t]
\caption{\em Basic information on \targfull. 
The $BV$ magnitudes are each the mean of 87 individual measurements \cite{Hog+00aa} distributed approximately randomly in orbital phase. 
The $JHK_s$ magnitudes are from 2MASS \cite{Cutri+03book} and were obtained at an orbital phase of 0.673. \label{tab:info}}
\centering
\begin{tabular}{lll}
{\em Property}                            & {\em Value}                 & {\em Reference}                      \\[3pt]
Right ascension (J2000)                   & 08 27 33.275                & \citenum{Gaia23aa}                   \\
Declination (J2000)                       & $-$20 50 38.25              & \citenum{Gaia23aa}                   \\
Bright Star Catalogue                     & HR 3335                     & \citenum{HoffleitJaschek91}          \\
Henry Draper designation                  & HD 71581                    & \citenum{CannonPickering19anhar}     \\
\textit{Gaia} DR3 designation             & 5706279565053294848         & \citenum{Gaia21aa}                   \\
\textit{Gaia} DR3 parallax                & $4.3083 \pm 0.1673$ mas     & \citenum{Gaia21aa}                   \\          
TESS\ Input Catalog designation           & TIC 144085463               & \citenum{Stassun+19aj}               \\
$B$ magnitude                             & $6.63 \pm 0.01$             & \citenum{Hog+00aa}                   \\          
$V$ magnitude                             & $6.59 \pm 0.01$             & \citenum{Hog+00aa}                   \\          
$J$ magnitude                             & $6.403 \pm 0.018$           & \citenum{Cutri+03book}               \\
$H$ magnitude                             & $6.410 \pm 0.024$           & \citenum{Cutri+03book}               \\
$K_s$ magnitude                           & $6.374 \pm 0.024$           & \citenum{Cutri+03book}               \\
Spectral type                             & A1~V + A1~V                 & \citenum{Andersen++84aa}             \\[3pt]
\end{tabular}
\end{table}

The variability of \targ\ was discovered from photographic patrol plates by Strohmeier, Knigge \& Ott \cite{Strohmeier++65ibvs2} without further comment. Andersen \& Nordstr\"om \cite{AndersenNordstrom77aas} found it to exhibit double lines which were sharp and underwent large radial velocity (RV) variations. Olsen \cite{Olsen77aas} obtained 67 photoelectric brightness measurements which showed it to have two sets of eclipses with approximately equal depth. The secondary eclipse occured at phase 0.48, indicating an eccentric orbit. It was given the name VV~Pyx in the \textit{65th Name-list of variable stars} \cite{Kholopov+81ibvs}. 

The only detailed analysis of \targ\ is by Andersen, Clausen \& Nordstr\"om \cite{Andersen++84aa} (hereafter ACN84), who obtained complete light curves (1495 points \cite{ClausenNordstrom80aa} in each of the $uvby$ passbands) using the Str\"omgren photometer \cite{Gronbech++76aas} on the Copenhagen 50~cm telescope at ESO La Silla, Chile, and photographic spectroscopy (28 high-dispersion photographic plates, each yielding an RV measurement for both stars). They confirmed the orbital eccentricity, detected apsidal motion, measured a spectral line strength ratio of $1.00 \pm 0.03$ between the stars, and found projected rotational velocities of $v \sin i = 23 \pm 3$\kms\ for both stars. Due to the similarity of the two stars, ACN84 assumed they were identical and presented the physical properties of the mean component of the system. A $V$-band light curve has since been presented by Shobbrook \cite{Shobbrook04jad}.

Samus et al.\ \cite{Samus+06astl} performed a comprehensive revision of the sky positions of objects included in the \textit{General Catalogue of Variable Stars} (GCVS\footnote{\texttt{http://www.sai.msu.su/gcvs/gcvs/}}). In 38 cases the variables were found to be in a different constellation than that adopted for their original GCVS designation, either due to changes in constellation boundaries, proper motion or improved measurement of their right ascension and declination. Kazarovets et al.\ \cite{Kazarovets+06ibvs} specified new names for these 38 variable stars, at which point our object of interest became \targ\ instead of VV~Pyx.

\targ\ has a close visual companion which is moderately fainter. The \textit{Index Catalogue of Visual Double Stars} \cite{Jeffers++63book} gives a separation of 0.3~arcsec and a brightness difference of 1.1~mag. Jens Viggo Clausen obtained a visual estimate of the magnitude difference of 2~mag on a night of good seeing, which is in good agreement with a spectroscopic measurements from two deep photographic plates (ACN84). ACN84 further suggested that the companion shows RV variability so could itself be a binary system. McAlister et al. \cite{McAlister+87aj,McAlister+90aj} found angular separations of 0.397~arcsec and 0.417~arcsec, respectively, via speckle interferometry; further measurements were made by this group but are not itemised here. The \gaia\ DR3 entry of \targ\ (Table~\ref{tab:info}) gives an unusually imprecise parallax ($4.31 \pm 0.17$~mas) and a large RUWE (renormalised unit weight error) of 4.8, suggesting the positional measurements were compromised by the nearby companion. The RUWE should be approximately 1.0, and a value above 1.4 is indicative of a poor astrometric solution \cite{Gaia21aa}; lower boundaries of 1.2 to 1.3 have been given by other authors \cite{Penoyre++22mn,Castroginard+24aa}.

ACN84 found an apsidal period of $U = 3200^{+1400}_{-800}$~yr. The relativistic contribution to this is significant, and it was identified by Gim\'enez \cite{Gimenez85apj} as a candidate for the detection of this phenomenon. The apsidal motion of the system has subsequently been discussed by many authors \cite{Claret97aa,PetrovaOrlov02ap,ClaretWillems02aa,Wolf+10aa}. The most recent work, by Claret et al.\ \cite{Claret+21aa} found a significantly shorter apsidal period of $758 \pm 29$~yr.

Finally, Kahraman Ali\c{c}avu\c{s} et al.\ \cite{Kahraman+23mn} included \targ\ in a list of 42 eclipsing systems which show pulsations. A set of peaks in the frequency spectrum of the system in the region of 35--40\cd\ were interpreted as resulting from $\delta$~Scuti pulsations.




\section*{Photometric observations}


\targ\ has been observed in three sectors by the NASA Transiting Exoplanet Survey Satellite \cite{Ricker+15jatis} (TESS). The data from sector 8 were obtained at a cadence of 1800~s, and from sectors 34 and 61 at a cadence of 120~s. Our analysis below concentrates on the data at 120~s cadence to avoid the smearing effects of the longer cadence. A fourth set of TESS observations (sector 88) was obtained in 2025 January but was not available when our analysis began.

\begin{figure}[t] \centering \includegraphics[width=\textwidth]{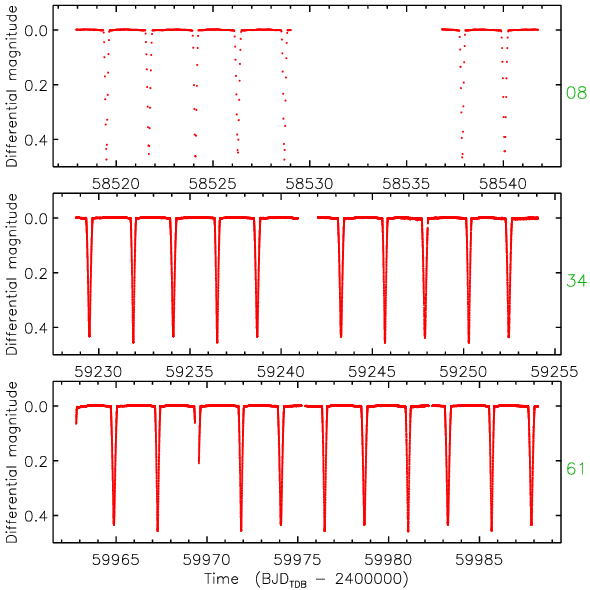} \\
\caption{\label{fig:time} TESS photometry of \targ. The flux measurements have been converted 
to magnitude units then rectified to zero magnitude by subtraction of the median. The data 
from sector 8 are from the QLP pipeline, and from sectors 34 and 61 from the SPOC pipeline. 
The sector number is shown in green to the right of each panel.} \end{figure}

The data were downloaded from the NASA Mikulski Archive for Space Telescopes (MAST\footnote{\texttt{https://mast.stsci.edu/portal/Mashup/Clients/Mast/Portal.html}}) using the {\sc lightkurve} package \cite{Lightkurve18}. We specified the quality flag ``hard'' to reject low-quality data, and used the simple aperture photometry (SAP) light curves from the SPOC data reduction pipeline \cite{Jenkins+16spie}. The data from sector 8 were only available in QLP (Quick-Look Pipeline) form \cite{Huang+20rnaas} and were also obtained using {\sc lightkurve}. We converted the datapoints into differential magnitude and subtracted the median magnitude from each sector for normalisation. The light curves are shown in Fig.~\ref{fig:time} and contain 747, 17373 and 17764 datapoints from sectors 8, 34 and 61, respectively.




\section*{Light curve analysis}

The eclipses in \targ\ occupy 14\% of each orbital period, and the remaining data hold little information about the properties of the system. We therefore extracted the data around each eclipse from the short-cadence light curves by retaining only the datapoints within one full eclipse duration of the midpoint of a fully-observed eclipse. This left a total of 4826 datapoints from sector 34, and 4753 from sector 61. We define star~A to be eclipsed at the deeper eclipse, and its companion to be star~B.

We modelled the light curves of the eclipses using version 43 of the {\sc jktebop}\footnote{\texttt{http://www.astro.keele.ac.uk/jkt/codes/jktebop.html}} code \cite{Me++04mn2,Me13aa}. The fitted parameters were the fractional radii of the stars ($r_{\rm A}$ and $r_{\rm B}$), expressed as their sum ($r_{\rm A}+r_{\rm B}$) and ratio ($k = {r_{\rm B}}/{r_{\rm A}}$), the central surface brightness ratio ($J$), third light ($L_3$), orbital inclination ($i$), orbital period ($P$), the reference time of primary minimum ($T_0$), and the quantities $e\cos\omega$ and $e\sin\omega$ where $e$ is the orbital eccentricity and $\omega$ is the argument of periastron. Limb darkening (LD) was accounted for using the power-2 law \cite{Hestroffer97aa,Maxted18aa,Me23obs2} with the same LD coefficients for both stars due to their similarity. The linear coefficient ($c$) was fitted and the non-linear coefficient ($\alpha$) held fixed at a theoretical value \cite{ClaretSouthworth22aa,ClaretSouthworth23aa}.

\begin{figure}[t] \centering \includegraphics[width=\textwidth]{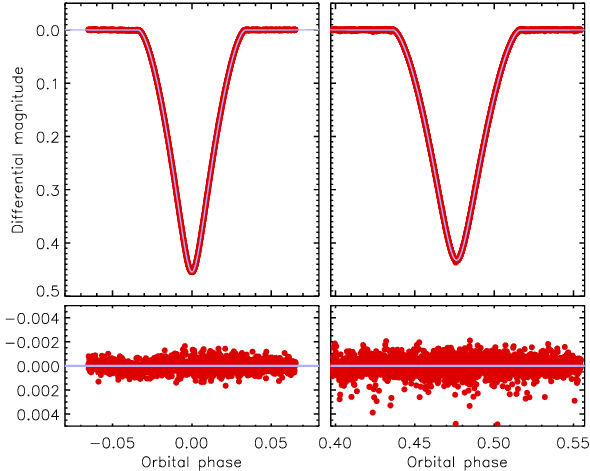} \\
\caption{\label{fig:phase34} {\sc jktebop} best fit to the light curves of \targ\ from 
TESS sector 34. The data are shown as filled red circles and the best fit as a light blue 
solid line. The residuals are shown on an enlarged scale in the lower panel.} \end{figure}

\begin{figure}[t] \centering \includegraphics[width=\textwidth]{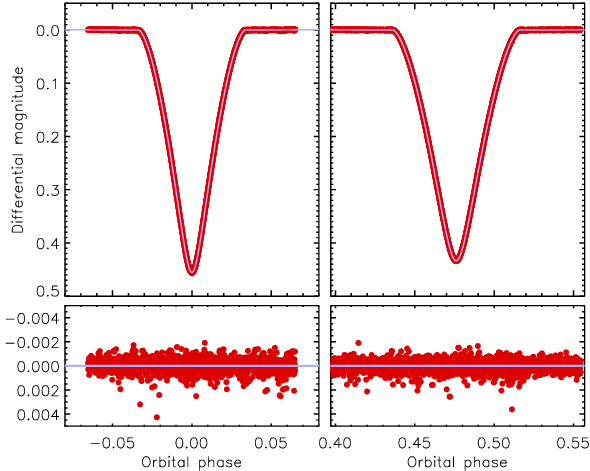} \\
\caption{\label{fig:phase61} {\sc jktebop} best fit to the light curves of \targ\ from 
TESS sector 61. Other comments are the same as for Fig.~\ref{fig:phase34}.} \end{figure}

Our initial fits to sectors 34 and 61 together led to a larger scatter than expected. We further constrained the ephemeris by measuring a time of primary eclipse from sector 8 and including it in the fit, finding that this made things worse (the best-fitting time of minimum from sector 8 was 29$\sigma$ from the measured value). From this we deduce that the effects of apsidal motion are significant over the time interval of the TESS observations, and also that the amount of third light may differ between sectors due to the different orientations of the TESS cameras. A natural solution is to model the data from sectors 34 and 61 individually, and leave discussion of the apsidal motion to future work. This resulted in much better fits being obtained, which are shown in Figs.\ \ref{fig:phase34} and \ref{fig:phase61}. The best-fitting parameter values are given in Table~\ref{tab:jktebop}. The residuals versus the best fits are larger in the secondary eclipse than in the primary eclipse, for both sectors. Inspection of the residuals versus time suggests that this is due to chance. We experimented with rejecting the eclipses with higher scatter but found that this had little effect on the results.

The parameter uncertainties were obtained using Monte-Carlo (MC) and resid-ual-permutation (RP) simulations ({\sc jktebop} tasks 8 and 9). As part of this process the data errors from the TESS reduction pipeline were normalised to force a reduced $\chi^2$ of unity. We were expecting the RP errorbars to be larger than the MC ones due to the pulsational signatures in the light curve, which were not included in the model thus are effectively red noise, but for most parameters the two were very similar. The main exception to this is $e\cos\omega$, from which we infer that the pulsations affect the measured phase difference between the eclipses.

\begin{sidewaystable} \centering
\caption{\em \label{tab:jktebop} Photometric parameters of \targ\ measured using 
{\sc jktebop} from the light curves from TESS sectors 34 and 61. The errorbars 
are 1$\sigma$ and were obtained from a residual-permutation analysis.}
\begin{tabular}{lccc}
{\em Parameter}                           &       {\em Value (sector 34)}      &       {\em Value (sector 61)}      &       {\em Adopted value}      \\[3pt]
{\it Fitted parameters:} \\                                                                                         
Orbital period (d)                        & $       4.5961598 \pm  0.0000025 $ & $       4.5961597 \pm  0.0000023 $ &                                    \\
Reference time (BJD$_{\rm TDB}$)          & $ 2459236.4944615 \pm  0.0000066 $ & $ 2459976.4770916 \pm  0.0000049 $ &                                    \\
Orbital inclination (\degr)               & $      88.1294    \pm  0.0055    $ & $      88.1387    \pm  0.0038    $ & $      88.1340    \pm  0.0055    $ \\
Sum of the fractional radii               & $       0.23018   \pm  0.00012   $ & $       0.23027   \pm  0.00011   $ & $       0.23022   \pm  0.00012   $ \\
Ratio of the radii                        & $       0.9820    \pm  0.0027    $ & $       0.9815    \pm  0.0021    $ & $       0.9817    \pm  0.0027    $ \\
Central surface brightness ratio          & $       0.99906   \pm  0.00035   $ & $       0.99944   \pm  0.00027   $ & $       0.99925   \pm  0.00035   $ \\
Third light                               & $       0.2026    \pm  0.0010    $ & $       0.1999    \pm  0.0010    $ &                                    \\
LD coefficient $c$                        & $       0.554     \pm  0.015     $ & $       0.560     \pm  0.011     $ & $       0.557     \pm  0.015     $ \\
LD coefficient $\alpha$                   &            0.4574 (fixed)          &            0.4574 (fixed)          &            0.4574 (fixed)          \\
$e\cos\omega$                             & $      -0.0370930 \pm  0.0000089 $ & $      -0.0374236 \pm  0.0000064 $ &                                    \\
$e\sin\omega$                             & $       0.08941   \pm  0.00021   $ & $       0.08963   \pm  0.00016   $ &                                    \\
{\it Derived parameters:} \\                                                                                                                            
Fractional radius of star~A               & $       0.11614   \pm  0.00017   $ & $       0.11621   \pm  0.00013   $ & $       0.11617   \pm  0.00017   $ \\
Fractional radius of star~B               & $       0.11404   \pm  0.00016   $ & $       0.11406   \pm  0.00013   $ & $       0.11405   \pm  0.00016   $ \\
Light ratio $\ell_{\rm B}/\ell_{\rm A}$   & $       0.9631    \pm  0.0051    $ & $       0.9626    \pm  0.0040    $ & $       0.9629    \pm  0.0051    $ \\[3pt]
Orbital eccentricity                      & $       0.09680   \pm  0.00019   $ & $       0.09713   \pm  0.00015   $ & $       0.09696   \pm  0.00019   $ \\
Argument of periastron ($^\circ$)         & $     112.533     \pm  0.048     $ & $     112.663     \pm  0.039     $ &                                    \\
\end{tabular}
\end{sidewaystable}


Table~\ref{tab:jktebop} contains the measured parameters from the fits to sectors 34 and 61, plus the larger of the MC or RP errorbar for each parameter. Given the similarity of the data and results for the two sectors we adopt the unweighted mean of those values in common. The uncertainties are extremely low so we take the largest of the four options for each parameter (Table~\ref{tab:jktebop}), thus forego the $\sqrt{2}$ boost from having two datasets fitted separately. 

The analysis above provides measurements of the fractional radii of the stars to a precision of 0.15\%. The surface brightness ratio is almost unity: the difference in the depths of primary and secondary eclipse is driven primarily by different fractions of the stars eclipsed at these times due to the eccentricity of the orbit. Two orbital ephemerides are also obtained, but are not valid for other time periods as they don't account for the apsidal motion.


\section*{Radial velocity analysis}

\begin{figure}[t] \centering \includegraphics[width=\textwidth]{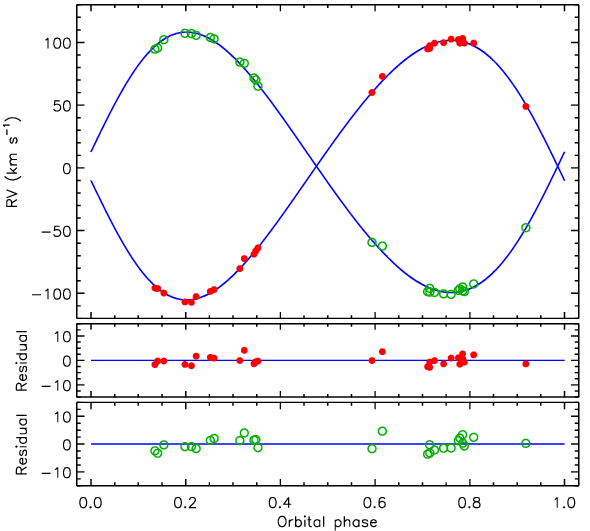} \\
\caption{\label{fig:rv} RVs of \targ\ from ACN84 compared to the best fit from 
{\sc jktebop} (solid blue lines). The RVs for star~A are shown with red filled 
circles, and for star~B with green open circles. The residuals are given in the lower panels 
separately for the two components.} \end{figure}

Armed with a new high-precision measurement of the orbital eccentricity, it was reasonable to reanalyse the published RVs of the stars (ACN84) to see if they were consistent and allowed a more precise measurement of the velocity amplitudes, $K_{\rm A}$ and $K_{\rm B}$. We manually digitised the RVs in table~2 of ACN84 then fitted spectroscopic orbits (see below). We found that the RVs for both stars at time HJD 2444348.5342 were highly discrepant, and that the time did not match the orbital phase in column 3 of the table. Changing the time to 2444384.5342 solved all three problems, and is more in line with the typical observing run time allocations\footnote{The times of the RVs are grouped into three discrete observing runs of 5--10 days each.}

We then fitted the RVs for the two stars using {\sc jktebop}. The only fitted parameters were $K_{\rm A}$, $K_{\rm B}$, $\omega$, the systemic velocities of the stars, and a phase offset versus our orbital ephemeris from TESS sector 34. The phase offset is small ($-0.011$), so our star identifications match those of ACN84. The systemic velocities of the stars were very similar so we ran a final fit with a value common to both stars. Errorbars were obtained from 1000 MC simulations \cite{Me21obs5}. The measured parameters are $K_{\rm A} = 103.47 \pm 0.37$\kms, $K_{\rm B} = 103.82 \pm 0.49$\kms, $\omega = 108.0^\circ \pm 2.0^\circ$ and a systemic velocity of $1.31 \pm 0.26$\kms\ (where the errorbar does not include uncertainty in the definition of the velocity scale). The fitted orbits are shown in Fig.~\ref{fig:rv}. The values and uncertainties we find are in good agreement with those from ACN84.


\section*{Physical properties and distance to \targ}

\begin{table} \centering
\caption{\em Physical properties of \targ\ defined using the nominal solar units 
given by IAU 2015 Resolution B3 (ref.~\cite{Prsa+16aj}). \label{tab:absdim}}
\begin{tabular}{lr@{\,$\pm$\,}lr@{\,$\pm$\,}l}
{\em Parameter}        & \multicolumn{2}{c}{\em Star A} & \multicolumn{2}{c}{\em Star B}    \\[3pt]
Mass ratio   $M_{\rm B}/M_{\rm A}$          & \multicolumn{4}{c}{$0.9966 \pm 0.0059$}       \\
Semimajor axis of relative orbit (\Rsunnom) & \multicolumn{4}{c}{$12.109 \pm 0.029$}        \\
Mass (\Msunnom)                             &  2.098  & 0.021       &  2.091  & 0.018       \\
Radius (\Rsunnom)                           &  2.1785 & 0.0072      &  2.1387 & 0.0070      \\
Surface gravity ($\log$[cgs])               &  4.0835 & 0.0024      &  4.0981 & 0.0020      \\
Density ($\!\!$\rhosun)                     &  0.2029 & 0.0011      &  0.2137 & 0.0011      \\
Synchronous rotational velocity ($\!\!$\kms)& 23.98   & 0.08        & 23.54   & 0.08        \\
Effective temperature (K)                   & 9500    & 200         & 9500    & 200         \\
Luminosity $\log(L/\Lsunnom)$               &  1.542  & 0.037       &  1.526  & 0.037       \\
$M_{\rm bol}$ (mag)                         &  0.885  & 0.092       &  0.925  & 0.092       \\
Interstellar reddening \EBV\ (mag)          & \multicolumn{4}{c}{$0.07 \pm 0.02$}			\\
Distance (pc)                               & \multicolumn{4}{c}{$178.4 \pm 2.5$}           \\[3pt]
\end{tabular}
\end{table}


We calculated the physical properties of \targ\ using the {\sc jktabsdim} code \cite{Me++05aa} with the $P$, $e$, $i$, $r_{\rm A}$ and $r_{\rm B}$ from our analysis of the TESS light curves, and the $K_{\rm A}$ and $K_{\rm B}$ from our analysis of the ACN84 RVs. We adopted an effective temperature of $\Teff = 8500 \pm 200$~K from ACN84 for both stars, as their surface brightnesses are practically identical. A somewhat larger value of $9311 \pm 195$~K was given by Zorec \& Royer \cite{ZorecRoyer12aa}. 

Our mass and radius measurements (Table~\ref{tab:absdim}) are consistent with those of ACN84, with the advantages that we have values for both stars rather than just the mean component, and that the radii are now measured to a precision of 0.3\% versus 0.9\%. The synchronous rotation velocities in Table~\ref{tab:absdim} are consistent with the measured values (ACN84).

The distance to \targ\ merits discussion. Inversion of the parallax from the \hip\ and \gaia\ satellites give distances of $222^{+63}_{-40}$~pc and $232 \pm 9$~pc, respectively. However, the \gaia\ value is unusually uncertain for a celestial object this close to the Solar system, and is accompanied by a RUWE indicating a poor astrometric fit (see \textit{Introduction}). ACN84 determined a distance of $195 \pm 10$~pc, somewhat shorter than both parallax-derived values (albeit that the \hip\ distance is very uncertain). 

We calculated a new distance estimate using the $BV$ and $JHK_s$ apparent magnitudes of the system (Table~\ref{tab:info}), the calibrations of surface brightness versus \Teff\ presented by Kervella et al.\ \cite{Kervella+04aa}, and the other quantities inputted to {\sc jktabsdim}. The $JHK_s$ magnitudes were converted to the Johnson system \cite{Carpenter01aj} but not corrected for the presence of the close companion. With an interstellar reddening of $\EBV = 0.07 \pm 0.02$~mag to equalise the distances at optical and infrared wavelengths, we obtained a much shorter distance of $162.8 \pm 2.2$~pc in the $K_s$ band. An alternative approach using theoretical bolometric corrections from Girardi et al.\ \cite{Girardi+02aa} gives a consistent distance of $165.1 \pm 2.3$~pc. 

The star within 0.4~arcsec of \targ\ will contaminate the photometry of the system. Without a precise magnitude difference and \Teff\ of the star we cannot properly correct for its contamination in the $BV $ and $JHK_s$ magnitudes. Its light acts to make \targ\ appear brighter and therefore closer to the observer. If we adopt a magnitude difference of 1.1~mag in all passbands \cite{Jeffers++63book} we find a distance of 190.1~pc instead of 162.8~pc; for a magnitude difference of 2~mag (ACN84) the distance becomes 175.3~pc. Our preferred option is instead to interpret the third light in our solutions of the TESS light curves as arising from the nearby star. In this case the amount of contamination is at least precisely determined, and we obtain a distance of 178.4~pc with a negligible additional uncertainty.


We cannot find a way to make our distance match that from \gaia\ DR3. Ignoring interstellar extinction changes the $K_s$-band distance by only +2.7~pc. Adding 1000~K to the \Teff\ values requires a larger \EBV\ and only shifts the distance measurement by +2.4~pc. Such a small effect might seem surprising, but is explicable: our preferred distance estimates rely primarily on the $K_s$ band, which is well into the Rayleigh-Jeans tail of the spectrum so is insensitive to temperature. The 2MASS $JHK_s$ apparent magnitudes were taken at orbital phase 0.673 so are well away from eclipse -- if they were in eclipse then this would make them fainter and bring the measured distance even closer. We conclude that the \hip\ parallax is too uncertain to conflict with our results, that the \gaia\ DR3 parallax is unreliable due to the close companion, and that our own distance measurements are also imperfect as they incorporate assumptions about the brightness and \Teff\ of the close companion.


\section*{Comparison with theoretical models}

\begin{figure}[t] \centering \includegraphics[width=\textwidth]{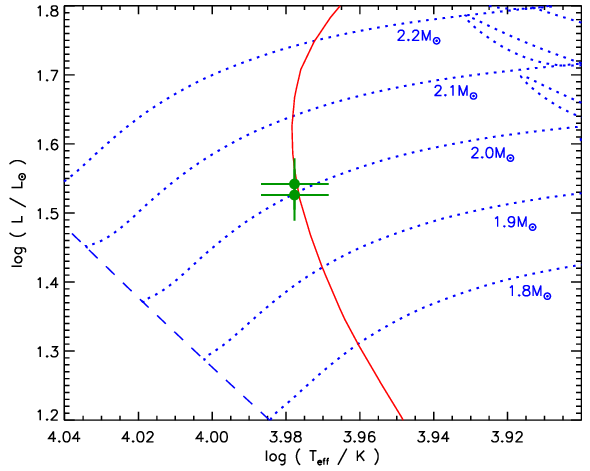} \\
\caption{\label{fig:hrd} Hertzsprung-Russell diagram for the components of \targ\ (filled 
green circles) and the predictions of the {\sc parsec} 1.2S models \cite{Chen+14mn} for 
masses of 1.8, 1.9, 2.0, 2.1 and 2.2\Msun\ (dotted blue lines with masses labelled) 
and the zero-age main sequence (dashed blue line), for a metal abundance of $Z=0.010$. 
The isochrone for an age of 570~Myr is shown with a solid red line.} \end{figure}

The similarity of the components of \targ\ means the system is not a particularly good test of theoretical models, but a comparison is still informative. We compared the measured properties of \targ\ to the predictions of the {\sc parsec} 1.2 theoretical stellar evolutionary models \cite{Bressan+12mn}, concentrating on the radii and \Teff\ values predicted for the known masses. 

A metal abundance of $Z=0.017$ and an age of 540~Myr fits the radii well but underpredicts the \Teff\ values by 600~K. A lower metal abundance of $Z=0.014$ requires an age of 560~Myr to match the radii but still underpredicts the \Teff\ values, by 400~K. Moving to a $Z$ of 0.010 and an age of $570 \pm 20$~Myr provides an excellent match to all three properties for both stars. This suggests that \targ\ is moderately metal-poor, something that should be confirmed spectroscopically. A Hertzsprung-Russell diagram is shown in Fig.~\ref{fig:hrd}.


\section*{Pulsation analysis}

\begin{sidewaysfigure} \centering
\includegraphics[width=\textwidth]{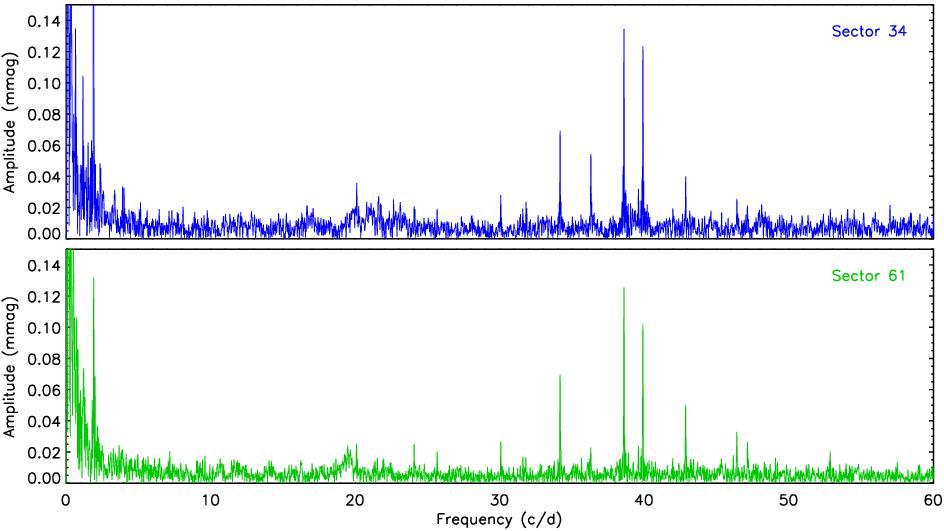}
\caption{\label{fig:freq} Frequency spectra of \targ\ from TESS sector 34 (upper panel, 
blue lines) and sector 61 (lower panel, green lines).} \end{sidewaysfigure}

The TESS light curve of \targ\ shows clear evidence for pulsations of relatively short period. The object has been previously identified as showing $\delta$~Scuti pulsations \cite{Kahraman+23mn}. We calculated frequency spectra for sectors 34 and 61 using version 1.2.0 of the {\sc period04} code \cite{LenzBreger05coast}. The sectors were treated individually to avoid problems with aliasing, and the {\sc jktebop} best fit was subtracted from the data prior to analysis. The frequency spectrum for sector 61 had a lower noise level so we used it to measure the significant frequencies in the light curve.

We found seven significant frequencies in the spectrum for sector 61, where we take the minimum signal-to-noise ratio (S/N) to be 4 (refs.\ \cite{Breger00aspc} and \cite{Kuschnig+97aa}). All of these are also present in the spectrum for sector 34, confirming their existence. An additional frequency at 36.3\cd\ is present in sector 61 but not sector 34 so may not be of astrophysical origin. The frequencies and their amplitudes are given in Table~\ref{tab:freq}, and the spectra are shown in Fig.~\ref{fig:freq}. 

Six of the frequencies are in the interval 30--50\cd\ so can be attributed to pulsations of the $\delta$~Scuti type. The \Teff\ and luminosity values of the stars put them slightly beyond the instability strip and in a region where the fraction of pulsators is approximately 0.1 (see Murphy et al.\ \cite{Murphy+19mn}). The remaining frequency is much lower, at 1.9\cd, and cannot be due to p-mode pulsations. It is instead borderline consistent with being of the SPB type \cite{Waelkens91aa}, with the components of \targ\ having \Teff\ values at the lower limit of this class \cite{Shi+23apjs}.

The orbital frequency of \targ\ is 0.2176\cd, and the Loumos \& Deeming \cite{LoumosDeeming78apss} frequency resolution is 0.10\cd. Frequencies $f_{3}$ and $f_{6}$ are close to being the 157th and 197th multiples of the orbital frequency, but this similarity is of low statistical significance given the frequency resolution of a single TESS sector. We are not able to attribute any frequency to an individual star with the available data, and indeed cannot rule out that some or all of the pulsations arise from the close companion.


\begin{table} \centering
\caption{\em Significant pulsation frequencies found in the TESS sector 61 light
curve of \targ\ after subtraction of the effects of binarity. \label{tab:freq}}
\setlength{\tabcolsep}{12pt}
\begin{tabular}{lccc}
{\em Label} & {\em Frequency (d$^{-1}$)} & {\em Amplitude (mmag)} & {\em S/N} \\[3pt]
$f_{1}$ & $ 1.9060 \pm 0.0010$ & $0.126 \pm 0.006$ &  5.0 \\
$f_{2}$ & $30.0686 \pm 0.0049$ & $0.027 \pm 0.006$ &  5.4 \\
$f_{3}$ & $34.1693 \pm 0.0019$ & $0.070 \pm 0.006$ &  9.2 \\  
$f_{4}$ & $38.5870 \pm 0.0011$ & $0.125 \pm 0.006$ & 11.8 \\
$f_{5}$ & $39.8933 \pm 0.0013$ & $0.102 \pm 0.006$ & 10.5 \\
$f_{6}$ & $42.8533 \pm 0.0027$ & $0.049 \pm 0.006$ &  6.5 \\  
$f_{7}$ & $46.3927 \pm 0.0040$ & $0.033 \pm 0.006$ &  4.8 \\
\end{tabular}
\end{table}


\section*{Summary and conclusions}

\targ\ is a dEB containing two A1~V stars in an orbit of period 4.596~d which shows both eccentricity and apsidal motion. We have determined the physical properties of the component stars using two sectors of short-cadence data from TESS and published photographic RVs from ACN84. We measure the radii of the stars individually for the first time, rather than the radius of the mean component of the system. The radii are extremely well-determined by the TESS data, and are consistent with the spectroscopic light ratio from ACN84. The properties of the system are best matched by theoretical predictions for stars of a metal abundance of $Z=0.010$ and an age of 570~Myr.

\targ\ has a companion at 0.4~arcsec which is fainter by 1.7~mag in the TESS passband, assuming it is the sole source of third light in the TESS data. This companion causes a poor fit to the astrometry in \gaia\ DR3, and thus an uncertain parallax. We instead measure a distance via the system's $K_s$-band apparent magnitude and calibrations of surface brightness versus \Teff, obtaining $178.4 \pm 2.5$~pc after correcting for the light from the third star under the assumption that it has the same \Teff\ as the eclipsing stars.

Pulsations are visible in the light curve of \targ. We subtracted the effects of binarity and measured seven significant pulsation frequencies in the data. Six of these are consistent with p-mode pulsations (30--46\cd) and one with g-mode oscillations (1.9\cd). We assign the higher frequencies to $\delta$~Scuti pulsations and the lower frequency to SPB pulsations; the component stars are outside but close to the instability strips for both types of variability. There is a chance that some or all of the pulsations arise from the fainter companion to the binary system.

The current work significantly increases the precision of the radius measurements of the members of the \targ\ system. Further improvements to the analysis could be obtained by better characterising the fainter nearby star, obtaining spectroscopic chemical abundances to check our inference of a low metallicity of the system, and precisely measuring its apsidal period to constrain the internal structure constants of the component stars


\section*{Acknowledgements}

This paper includes data collected by the TESS\ mission and obtained from the MAST data archive at the Space Telescope Science Institute (STScI). Funding for the TESS\ mission is provided by the NASA's Science Mission Directorate. STScI is operated by the Association of Universities for Research in Astronomy, Inc., under NASA contract NAS 5–26555.
This work has made use of data from the European Space Agency (ESA) mission {\it Gaia}\footnote{\texttt{https://www.cosmos.esa.int/gaia}}, processed by the {\it Gaia} Data Processing and Analysis Consortium (DPAC\footnote{\texttt{https://www.cosmos.esa.int/web/gaia/dpac/consortium}}). Funding for the DPAC has been provided by national institutions, in particular the institutions participating in the {\it Gaia} Multilateral Agreement.
The following resources were used in the course of this work: the NASA Astrophysics Data System; the SIMBAD database operated at CDS, Strasbourg, France; and the ar$\chi$iv scientific paper preprint service operated by Cornell University.



\end{document}